\documentclass[prl,twocolumn,letterpaper,amsmath,amssymb,amsfonts,superscriptaddress,showpacs]{revtex4}

\usepackage{graphicx}
\usepackage{epstopdf}
\usepackage{amsmath}
\usepackage{braket}
\usepackage{color}
\usepackage{subfigure}
\usepackage{graphics}
\usepackage{harpoon}
\usepackage{multirow}
\usepackage{pdfpages}

\definecolor{nblue}{rgb}{0.3,0.3,1.0}
\definecolor{ngreen}{rgb}{0.2,0.7,0.2}
\definecolor{nred}{rgb}{1.0,0.2,0.2}
\definecolor{nyellow}{rgb}{0.9,0.7,0.2}
\definecolor{npurple}{rgb}{0.8,0.2,0.8}
\definecolor{nrose}{rgb}{0.5,0.3,0.3}
\definecolor{nbackground}{rgb}{1,1,1}
\definecolor{ngrey}{rgb}{0.5,0.5,0.5}
\definecolor{nbrown}{rgb}{0.5,0.4,0.0}
\definecolor{nblack}{rgb}{0,0,0}
\definecolor{ncyan}{rgb}{0.1,0.5,0.5}
\definecolor{npale}{rgb}{0.9,0.9,0.9}

\newcommand{\dd}{\mathrm{d}}

\newcommand{\bku}{\color{nblue}}

\newcommand{\beq}{\begin{equation}}
\newcommand{\eeq}{\end{equation}}

\newcommand{\erf}[1]{Eq.~(\ref{#1})}
\newcommand{\erfs}[2]{Eqs.~(\ref{#1})--(\ref{#2})}

\newcommand{\cu}[1]{\left\{ {#1} \right\}}

\newcommand{\Tr}{\text{Tr}}

\newcommand{\s}[1]{\hat{\sigma}_{#1}}

\newcommand{\past}[1]{\overleftarrow{#1}}
\newcommand{\fut}[1]{\overrightarrow{#1}} 
\newcommand{\both}[1]{\overleftrightarrow{#1}}
\newcommand{\retro}{_{\text R}}
\newcommand{\fil}{_{\text F}}
\newcommand{\sm}{_{\text S}}
\newcommand{\god}{_{\text T}}

\renewcommand{\section}[1]{{\em #1}.---}
\newcommand{\letter}{Letter}
\newcommand{\ti}{ t_0 }

\newcommand{\nbf}[1]{ {#1} }
\newcommand{\con}{_{\rm C}}
\newcommand{\blu}{\bku}
\renewcommand{\god}{_{\rm T}}

\begin{document}

\title{Quantum State Smoothing}

\author{Ivonne Guevara}
\affiliation{Centre for Quantum Computation and Communication Technology (Australian Research Council), Centre for Quantum Dynamics, Griffith University, Brisbane, Queensland 4111, Australia}
\author{Howard Wiseman}
\affiliation{Centre for Quantum Computation and Communication Technology (Australian Research Council), Centre for Quantum Dynamics, Griffith University, Brisbane, Queensland 4111, Australia}

\date{\today}

\begin{abstract}
Smoothing is an estimation method whereby a classical state (probability distribution for classical variables) at a given time is conditioned on all-time (both earlier and later) observations. Here we define a smoothed quantum state for a partially monitored open quantum system, conditioned on an all-time monitoring-derived record. We calculate the smoothed distribution for a hypothetical unobserved record which, when added to the real record, would complete the monitoring, yielding a pure-state ``quantum trajectory''. Averaging the pure state over this smoothed distribution yields the (mixed) smoothed  quantum state.  
We study how the choice of actual unravelling affects the purity increase over that of the conventional (filtered) state conditioned only on the past record. 

\end{abstract}
\maketitle

Estimation theory is used to assign values to parameters of interest, whose true values are unknown, using   the available data. These parameters may evolve dynamically, and new data may arrive dynamically, through a continuous measurement process. Estimation in this instance is non trivial because there can be noise associated with the measurement, noise affecting directly the dynamical system due to its environment, and initial uncertainty in the parameters. Optimal estimation theory can be formulated using the Bayesian approach to statistics, whereby the observer's knowledge of the parameters is described by a conditional probability distribution  $\wp_{\rm C}$, also known as a Bayesian state. If this state is conditional on measurements at earlier times, it is called a {\em filtered} state $\wp_{\rm F}$, while if it is conditional on all-time (both earlier and later)  measurements,  it is called a {\em smoothed} state $\wp_{\rm S}$. Smoothing uses more complete information than filtering, and so typically delivers
a probability distribution that is purer (that is, having less entropy). 

In the  flourishing and  important area of {\em quantum} estimation theory~\cite{WM10,Hel76,GLM04,GLM11,Hol12,BraKha92,TWC11,TN12,Tsa13,Tsa09c,TJ&13,BI&96},  much has been learnt from classical estimation theory. The analogy between quantum states and classical Bayesian states 
has been fruitful even in quantum foundations~\cite{GarZol04,CavFucSch02a,FucMerSch14}. In particular, the stochastically evolving conditioned state $ \rho_{\rm C}$ of an open quantum state, as introduced by physicists~\cite{Car93b,DalCasMol92,GarParZol92,WisMil93c,Bar93,KorPRB99} and applied in quantum control~\cite{WisMil93b,Wis95c,HorKil97,Doh00,Smi02,WisDoh05,Bla06,WM10,VM&12}, is now understood to be analogous to the classical filtered state~\cite{Bel88,Bel99,Doh00,Jam04,HofMahHes98,KorPRB01b}  $\wp_{\rm F}$ 
(and so for clarity we write it as $\rho_{\rm F}$).  However, the situation is very different regarding smoothing. 
 
The term ``quantum smoothing'' was introduced by Tsang~\cite{Tsa09prl,Tsa09praI}  in 2009  
to mean smoothed estimation of   classical parameters that affect the evolution of a quantum system, 
 using the results of measurements on that system.   It has been shown to be useful to the problem of estimating a stochastically 
varying optical phase using the all-time photocurrent record, both theoretically~\cite{TsaShaLlo09,BerHalWis13}  and experimentally~\cite{Whe&10,YH&12}. It has also been applied to the problem of estimating an unknown result from a measurement on a quantum system at one time, using records obtained both before and after that time, again both theoretically~\cite{GJM13} and experimentally~\cite{TWSea14}. 
However, none of the above define a quantum smoothed state---that is,  
a positive operator $\rho_{\rm S}$ that is analogous to a Bayesian smoothed state $\wp_{\rm S}$. 
There is a good reason for this lack, which is most easily stated in the Heisenberg picture~\cite{Tsa09prl}: 
quantum operators for a system at a given time commute with operators representing the results of earlier measurements on that system, but do not commute with operators representing the results of later measurements on that system~\cite{WM10}. 


In this \letter\ we show that there is a situation in which it {\em is} possible to define 
\textit{quantum state smoothing},  producing a positive state $\rho_{\rm S}$ conditioned on 
both earlier and later results. 
The situation is that of open quantum systems under partial observation, 
which is the typical situation experimentally~\cite{Smi02,Bla06,VM&12}. The system has couplings to 
several baths (all assumed Markovian). An experimenter, Alice, 
is able to monitor some of them, yielding the record $ \mathbf{O}$ she observes.  
Other baths are not monitored by her, but hypothetically they could be monitored by another party, yielding results $ \mathbf{U}$ unobserved by Alice. The `true' state $\rho_{\rm T}$ 
conditioned on both observed ($ \mathbf{O}$) and unobserved ($ \mathbf{U}$) records 
would be pure, while that conditioned only on $ \mathbf{O}$ is mixed. 
The crucial point is that the record $\mathbf{U}$, comprising classical variables ($c$-numbers), can be estimated 
by applying smoothing to the record $ \mathbf{O}$, and in this way Alice can obtain a {\em smoothed quantum state} 
$\rho_{\rm S}$. As in the classical case, this is typically purer than $\rho_{\rm F}$, and a better approximation to $\rho_{\rm T}$. 


We first review the necessary theoretical background. 
Then we explain how our smoothed quantum state $\rho_{\rm S}$ is 
quite different from the ``past quantum state'' (actually a pair of operators) introduced in 
Ref.~\cite{GJM13}. We also  
show that our approach subsumes classical state estimation by smoothing, 
also known as the hidden Markov model (HMM) technique~\cite{PTVF07}. A HMM is applicable to quantum systems only when they are effectively classical (i.e.~always diagonal in a particular basis)  as in Refs.~\cite{Arm09,GMKM14, RPP15}. We apply our method to a genuinely quantum system (i.e.~one which is not diagonal in a fixed basis)---a coherently driven two-level atom, the radiation from which is partly observed. We take the known record to be generated by homodyne detection, and the unknown record to be that corresponding to photon absorption, as this is the most intuitive picture of what happens to photons that are lost into the laboratory surroundings. These lost photons result in impurity in the standard (filtered) conditioned system state. We show that our smoothing technique can, on average, eliminate up to  26\%   of this impurity. 
Our investigations shed light on how well we can know the trajectory of a partially observed open quantum systems, and the relation between the quantum and classical versions of state smoothing. 

\section{Types of Estimation} 
 Consider a  classical  dynamical system described by parameters ${\bf x}_t$ (bold font indicates a vector of parameters)  which is monitored to yield a noisy output at each time, ${\bf r}_t$. 
We denote a {\em measurement record} ${\bf R}_{\Omega} = \cu{{\bf r}_t\,:  t \in \Omega}$, where $\Omega\subseteq[\ti,T]$ 
is typically some 
finite time interval. Bayesian estimation involves data processing to infer the conditional  classical state 
\begin{equation}
  \wp_{{\bf R}_\Omega}({\bf x}_{\tau})\equiv \Pr[  {\bf x}^{\rm true}_\tau={\bf x}_\tau|{\bf R}_\Omega ; \wp_0] ,
\end{equation}
where $\wp_0$ describes the {\em a priori} statistics of ${\bf x}$ at the initial time $\ti$. It is also useful to define the 
unnormalized state
\beq  \label{eq:unnormclasstate}
\tilde\wp_{{\bf R}_\Omega}({\bf  x}_{\tau})\equiv \wp_{{\bf R}_\Omega}({\bf x}_{\tau}) 
\frac{\wp({\bf R}_\Omega|\wp_0)}{\wp_{\rm ost}({\bf R}_\Omega|\wp_0)} \propto \wp({\bf R}_\Omega,{\bf  x}_{\tau}|\wp_0).
\eeq 
Here the  $\wp({\bf R}_\Omega|\wp_0)$ is the  actual  distribution for ${\bf R}_\Omega$ while 
$\wp_{\rm ost}({\bf R}_\Omega|\wp_0)$ is an 
`ostensible' distribution for it---it is positive and normalized, but 
 is otherwise arbitrary and does not depend on ${\bf x}_t$ \cite{GueWis15}. 

\begin{figure}[ht]
\begin{center}
\includegraphics[scale=.32]{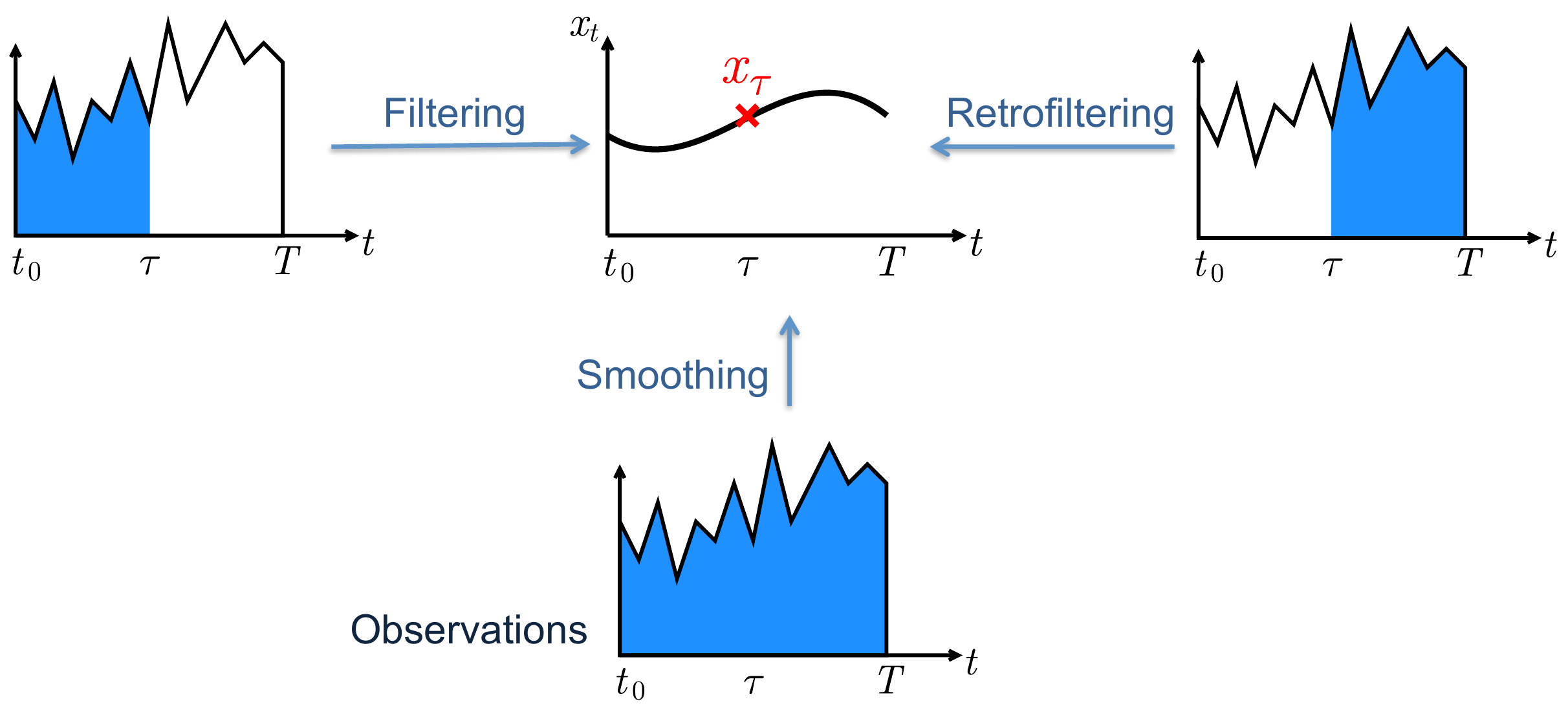}
\caption{Classical estimation classes depending on the measurement record considered relative to $\tau$, time at which the signal is to be estimated. (Adapted from \cite{Tsa09prl}).}
\label{fig:Estimation}
\end{center}
\end{figure}

There are three types of estimation worth distinguishing~\cite{Par81,Jun11}: \textit{filtering}, \textit{retro-filtering} (as we call it), and \textit{smoothing} (see Fig.~\ref{fig:Estimation}). 
If---as in feedback control problems---for  the time of interest $\tau$ there is only access to earlier results,   $\past{\bf R}_{ \tau} \equiv {\bf R}_{[\ti,\tau)}$, the optimal protocol is filtering: $\wp\fil({\bf x}_{\tau}) \equiv \wp_{\past{\bf R}_{\blu\tau}}({\bf x}_{\tau})$. If there is access only to 
later results, $\fut{\bf R}_{ \tau} \equiv {\bf R}_{[\tau,T)}$,  the optimal protocol is retro-filtering: $\wp\retro({\bf  x}_{\tau}) \equiv \wp_{\fut{\bf R}_{\blu \tau}}({\bf  x}_{\tau})$.   As its name implies, this is simply the time-reverse to filtering, but starting with an uninformative final state $\wp({\bf  x}_{T}) \propto 1$. Finally, if  the all-time  record $\both{\bf R} \equiv {\bf R}_{[\ti,T)}$ is available, 
 with $\ti<\tau<T$, 
then all the information can be utilised,  by the technique of  smoothing: $\wp\sm({\bf  x}_{\tau}) = \wp_{\both{\bf R}}({\bf  x}_{\tau})$.  This combines  filtering and retrofiltering, using unnormalized states~\cite{Tsa09praI}: 
\begin{equation}\label{eq:classSmooth}
\wp\sm({\bf  x}_{\tau})=\frac{\tilde\wp\retro({\bf  x}_{\tau}) \tilde\wp\fil({\bf  x}_{\tau})}{\int\dd {\bf  x}'_{\tau} \tilde\wp\retro({\bf  x}'_{\tau}) \tilde\wp\fil({\bf  x}'_{\tau}) }.
\end{equation}
Here one of the states (most conveniently the retrofiltered one) is defined using an uninformative prior, $\wp_0 \propto 1$, to prevent double counting of the {\em a priori} information.

\section{Quantum analogues of (retro)filtering}
An extension of these results to quantum mechanics has been done partially. Quantum trajectory theory~\cite{Car93b} is  the analogue of classical state filtering.  A quantum trajectory describes the path (in ``density operator space'') of the state of the quantum system through time, conditioned on  the measurement result ${\bf r}_{t}$ in each infinitesimal interval  $[t , t+\dd t)$.  Note that the path may be continuous (quantum diffusion) or discontinuous (quantum jumps)~\cite{Car93b,WM10}.  
This process is described by a set 
(indexed by ${\bf r}_{t}$) of measurement operations  (completely positive maps)  ${\cal M}_{{\bf r}_{t}}$, that evolve the state 
forward in time~\cite{GueWis15}:
 \begin{equation}\label{eq:rhofil}
\tilde\rho\fil(t+ \dd t)  
 ={\cal M}_{{\bf r}_{t}}\,\tilde\rho\fil(t).
\end{equation} 
Starting with $\rho(t_0)=\rho_0$, this procedure generates the state conditioned on the whole past record: $\tilde\rho\fil(\tau) = \tilde\rho_{\past{\bf R}_{\tau}}({\tau})$. 
This is an unnormalized state (as indicated by the tilde), analogous to \erf{eq:unnormclasstate}. That is, the normalized version $\rho\fil(\tau)$ generates the correct filtered probability distribution $\wp\fil({ x}_\tau)$ for any system observable $\hat{X}_\tau$, while 
\begin{equation} \label{eq:and5}
\Tr[\tilde\rho\fil
({\tau})] {\wp_{\rm ost}(\past{{\bf R}}_{\tau}|\rho_0)} =  { \wp(\past{\bf R}_{ \tau}|\rho_0)} .
\end{equation}

The corresponding analogue for Bayesian state retrofiltering has been  set out in \cite{Tsa09prl};  it 
is the solution of the adjoint of equation \eqref{eq:rhofil}, 
\begin{equation}\label{eq:Eretro}
\hat{E}\retro(t) 
={\cal M}_{{\bf r}_{t}}^{\dagger}\, \hat{E}\retro(t+\dd t).
\end{equation}
In this case the \emph{effect operator} evolves backwards from the final uninformative effect $\hat{E}(T)=I$ towards $\hat E\retro(\tau) \equiv \hat E_{\fut{\bf R}_{\tau}}(\tau)$,  conditioned on the record $\fut{\bf R}_{\tau}$ in the future of $\tau$. This solution $\hat E_{\fut{\bf R}_{\tau}}(\tau)$ determines the statistics of $\fut{\bf R}_{\tau}$: 
\begin{equation}\label{eq:TrEyrho}
\Tr[\hat E\retro
(\tau){\rho}_\tau]\wp_{\rm ost}(\fut{\bf R}_{\tau}|\rho_\tau) = \wp(\fut{\bf R}_{\tau}|\rho_\tau).
\end{equation}

\textit{Quantum smoothing?---} 
A naive approach to construct a quantum smoothed state, given the quantum analogues of filtering $\rho\fil$ and retrofiltering $\hat{E}\retro$, would be to combine them directly  as  in equation Eq.~\eqref{eq:classSmooth} so that $\rho\sm(\tau)\propto\rho_{\text{F}}({\tau})\hat{E}\retro(\tau)$. However, as pointed out in Ref.~\cite{Tsa09prl}, the result is not in general Hermitian or (even if symmetrized) positive semi-definite.  Therefore it cannot correspond to a physical state.  
 As discussed in the introduction, there is a deep reason for this, which is 
why Tsang gave quantum smoothing the restricted meaning of estimating an external {\em classical} parameter ${\bf x}$. 

The filtered state $\rho\fil(\tau)$ and the retrofiltered effect $\hat{E}\retro(\tau)$ are sufficient to best estimate, from the all-time record $\both{\bf R}$, the result of any measurement performed on the system at time $\tau$. For this reason,  Gammelmark {\em et al.} 
declared the pair $\Xi=(\rho\fil,\hat E\retro)$ to be ``the past quantum state''~\cite{GJM13}. 
They did not define a quantum state in the usual sense ({\em i.e.}~a density operator) combining the future and past information. 
While our notion of quantum state smoothing also makes use of 
filtered states and retrofiltered effects, it is quite different in that: it applies only to partially observed open quantum systems; 
 it defines a quantum state in the usual sense; and it can be compared directly to quantum state filtering by measures such as purity, similarly to the classical case. 


\begin{figure}[t]
\begin{center}
\includegraphics[scale=.35]{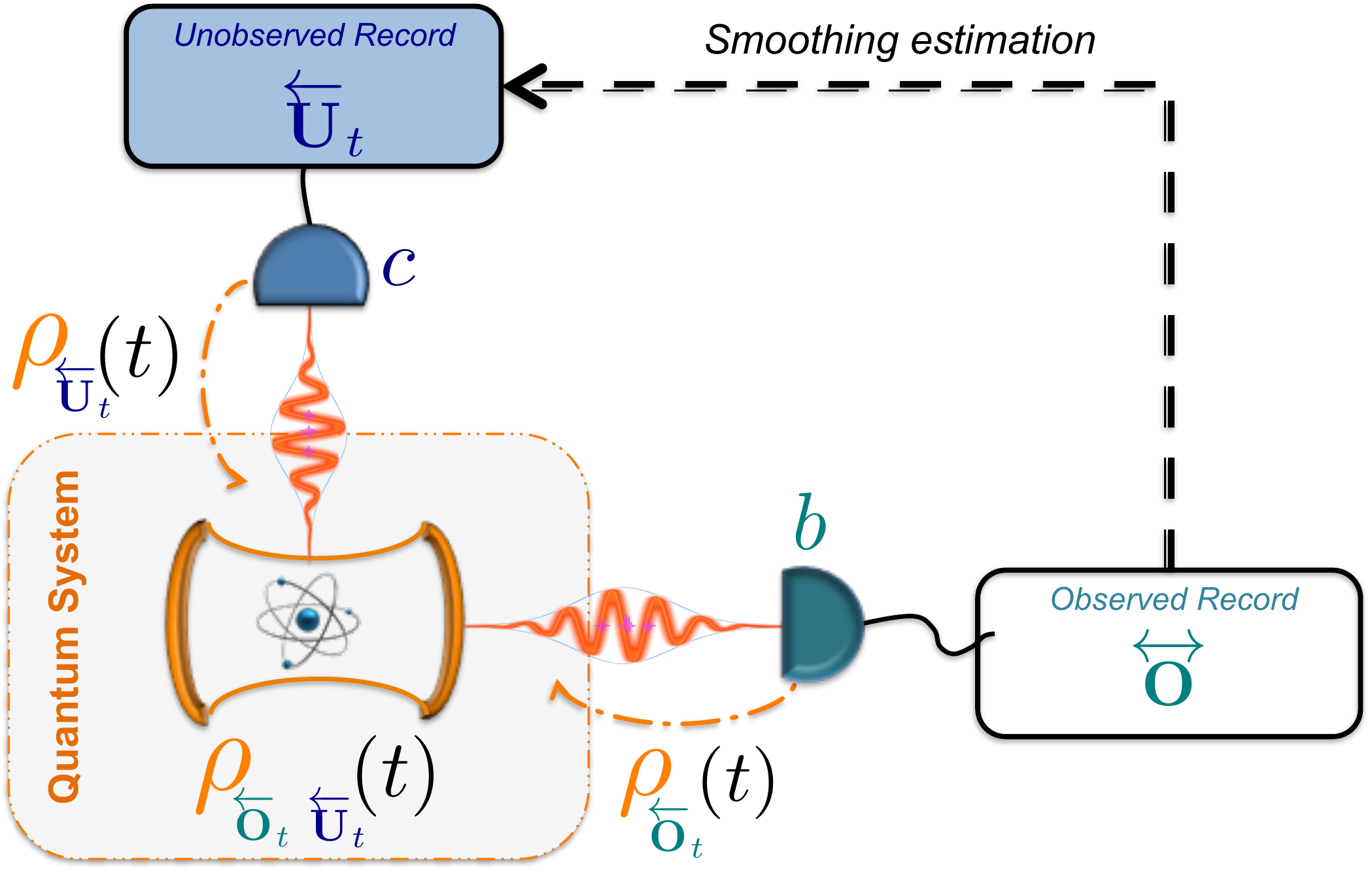}
\caption{The quantum state smoothing problem:  to best approximate the unknown true state of a quantum system, conditioned on both observed ( $\bf O$) and unobserved ( $\bf U$) records, given access only to  $\bf O$. This requires one to estimate 
 \unexpanded{$\past{\bf U}_t$}  up to time $t$ using the full record for  \unexpanded{$\both{\bf O}$}  (before and after $t$).}
\label{fig:QT}
\end{center}
\end{figure}

\section{Quantum state smoothing} 
The key idea for quantum state smoothing is illustrated in Fig.~\ref{fig:QT}. 
Consider an open quantum system with two  groups of  output channels ($b$, $c$).  
An observer Alice monitors the first group, $b$, yielding the all-time 
measurement record $\both{\bf O}$.  A hypothetical observer monitors the second group $c$, 
yielding a record $\both{\bf U}$ unobserved by Alice.  
The `true' state of the system $\rho_{\rm T}(t) \equiv \rho_{\past{\bf O}_{t},\past{\bf U}_{t}}(t)$  is conditioned on   both measurement records.  If $\rho_0$ is pure then $\rho_{\rm T}(t)$ will be pure for all times; no extra conditioning could possibly give a better (more pure) state. However Alice's conditional state, 
calculated in the conventional way (filtering),  
\beq
\rho\fil(t) \equiv \rho_{\past{\bf O}_{t}}(t) = {\rm E}_{\past{\bf U}_{t}|\past{\bf O}_t} [\rho_{\past{\bf O}_{t},\past{\bf U}_{t}}(t) ],
\eeq
will be mixed, because of the averaging over the unobserved record---here ${\rm E}_{A|B}[X]$ means the expected value of $X$, averaged over $A$, for a given $B$.  Note that this averaging does not have to be done explicitly---it is implicit in the quantum trajectory theory as in \erf{eq:rhofil} and is independent of how the channels $c$ are monitored. 

 The crucial insight is that Alice can do better, when averaging over $\past{\bf U}_t$, by using information 
  in the future of $t$, to define a positive-definite smoothed quantum state 
  \begin{equation}\label{eq:defsqs}
\rho\sm(t) =  {\rm E}_{\past{\bf U}_{t}|\both{\bf O}} [\rho_{\past{\bf O}_{t},\past{\bf U}_{t}}(t) ] \equiv \sum_{\past{\bf U}_{t}} \wp\sm(\past{\bf U}_{t})  \rho_{\past{\bf O}_{t},\past{\bf U}_{t}}(t), 
\end{equation}
Here $\wp\sm(\past{\bf U}_{t})= \wp_{\both{\bf O}}(\past{\bf U}_{t}) 
=\Pr[\past{\bf U}_{t}^{\rm true}=\past{\bf U}_{t}|\both{\bf O},\rho_0]$ is the probability distribution for the unobserved record prior to $t$, obtained by smoothing from  the all-time observed record $\both{\bf O}$. Note that $\wp\sm(\past{\bf U}_{t})$ yields exactly the same type of information as the 
``past quantum state'' of Ref.~\cite{GJM13}, except that it is more general --- it specifies 
the probability of a continuous 
monitoring record $\past{\bf U}$, not just a result of a measurement at 
one point in time, and is conditioned on another record, $\both{\bf O}$, 
covering the same time interval, not 
just records strictly earlier and strictly later.  Unlike the classical stochastic process ${\bf x}_t$ considered previously by Tsang~\cite{Tsa09prl}, the record $\past{\bf U}_{t}$ is of 
quantum origin---its statistics are undefined without a quantum system. Nevertheless, it is still a time-series of $c$-numbers with well-defined statistics and so there is no conceptual problem in applying his theory of quantum smoothing to obtain $\wp\sm(\past{\bf U}_{t})$, and thereby $\rho\sm(t)$.


We now show how to calculate \erf{eq:defsqs}.  For definiteness and simplicity,  we consider a single channel ($b$) yielding homodyne photocurrent $y_t$ and a single channel ($c$) yielding an unobserved photon count $n_t$.  These processes are 
related to the dynamics of the quantum system via the 
joint measurement operation ${\cal M}_{{\nbf n}_t,{\nbf y}_t}$ defined such that ${\cal M}_{{\nbf {y}}_t} = \sum_{{\nbf{n}}_t=0}^1  \wp_{\rm ost}({\nbf{n}}_t|\past{\nbf{Y}_{ t}}){\cal M}_{{\nbf n}_t,{\nbf y}_t}$, 
for a convenient choice of $\wp_{\rm ost}({\nbf{n}}_t|\past{\nbf{Y}_{ t}})$~\cite{GueWis15}. 
 By standard techniques~\cite{WM10}, ${\cal M}_{{\nbf n}_t,{\nbf y}_t}$  lets us  generate a typical sample of the all-time records $\both{\nbf{O}}=\both{\nbf{Y}}$ and $\both{U}=\both{N}^{\rm true}$. For all but one purpose (see below), the latter is irrelevant, but using the former we  calculate the filtered state  $\rho_{\past{\nbf{Y}_{ t}}}(t)$ from Eq.~\eqref{eq:rhofil} and  the 
retrofiltered effect 
$\hat{E}_{\fut{\nbf{Y}_{t}}}(t)$ from  Eq.~\eqref{eq:Eretro}, 
 with ${\cal M}_{y_t}$ in place of ${\cal M}_{{\bf {r}}_t}$~\footnote{We tested the correctness of our numerical calculations for the  two-level atom  example below  by  verifying the $t$-independence of \unexpanded{$\wp(\both{\nbf{Y}})=\Tr[\hat{E}_{\fut{\nbf{Y}_{t}}}(t)\rho_{\past{\nbf{Y}_{t}}}(t)]$}.}.  
 Next, we generate a large ensemble $ \mathfrak{E}_{\rm ost}$ of random samples of  $\both{\nbf{U}}=\both{\nbf N}$, according to the ostensible distribution $\wp_{\text{ost}}({\nbf n}_t|\past{\nbf{Y}_{ t}})$. For each sample we calculate an associated  
pure state $\tilde{\rho}_{\past{\nbf Y_{t}},\past{\nbf N}_{ t}}(t)$, conditioned on both 
records, from Eq.~\eqref{eq:rhofil} with ${\cal M}_{n_t,y_t}$ in place of 
${\cal M}_{{\bf  {r}}_t}$~\footnote{We tested this numerically for the example below by verifying that the ensemble average of the calculated `doubly' filtered states over the random sample of \unexpanded{$\both {\nbf N}$} coincides with the filtered states only conditioned on \unexpanded{$\past{\nbf Y}_{t}$} i.e., \unexpanded{${{\rm E}_{\past{\nbf N_{t}}}}[{\rho_{\past {\nbf N_{t}},\past{\nbf Y_{t}}}(t)}]={\rho_{\past{\nbf Y_{t}}}(t)},\,\forall t\in[\ti,T]$}. }.

Elementary manipulation of probabilities~\cite{GueWis15} gives $\wp\sm(\past{\nbf N_{ t}}) \equiv  
\wp(\past{N}_t|\both{Y})
\propto \wp(\fut{\nbf Y_{ t}}|\past{\nbf N_{ t}},\past{\nbf Y_{ t}})\,\wp(\past{\nbf N_{ t}}|\past{\nbf Y_{t}})$. 
Using the  equations for multiple channels corresponding to  Eq.~\eqref{eq:TrEyrho},
\begin{equation}\label{eq:py|ny}
\begin{array}{rl}
\wp(\fut{\nbf Y_{ t}}|\past{\nbf N_{ t}},\past{\nbf Y_{ t}})&=\Tr[\hat{E}_{\fut{\nbf Y_{ t}}}\rho_{\past{\nbf N_{ t}}\past{\nbf Y_{ t}}}]\,\wp_{\text{ost}}(\fut{Y_{ t}}), 
\end{array}
\end{equation}
and to  \erf{eq:and5},  
\begin{equation}\label{eq:TrErhoun}
\Tr[\hat{E}_{\fut{\nbf Y_{ t}}}\tilde{\rho}_{\past{\nbf N_{ t}}\past{\nbf Y_{ t}}}]\,\wp_{\text{ost}}(\past{\nbf N_{ t}}|\past{\nbf Y_{ t}	})=\Tr[\hat{E}_{\fut{\nbf Y_{ t}}}\rho_{\past{\nbf N_{ t}}\past{\nbf Y_{ t}}}]\,\wp(\past{\nbf N_{ t}}|\past{\nbf Y_{ t}}), 
\end{equation}
we finally obtain, from \erf{eq:defsqs}, 
\begin{equation}
\rho\sm(t) 
\propto \sum_{\past{\nbf N_{ t}}}\wp_{\text{ost}}(\past{\nbf N_{ t}}|\past{\nbf Y_{t}})
\times 
{\rho}_{\past{\nbf Y_{ t}},\past{\nbf N_{ t}}}(t) \,\Tr[\hat{E}_{\fut{\nbf Y_{ t}}}(t) \tilde{\rho}_{\past{\nbf Y_{ t}},\past{\nbf N_{ t}}}(t)] .  
\end{equation} 
 We can approximate this weighted average over all possible unobserved records 
using the ensemble $ \mathfrak{E}_{\rm ost}$ drawn from the appropriate ostensible distribution,  as 
discussed in the preceding paragraph. 
 This is the method we use below to find the smoothed quantum state. 

\section{Example} 
Consider  a two-level atom,  with driving Hamiltonian $\hat H = (\Omega/2) \s{x}$ in the interaction frame,  
and radiative damping described by  a Lindblad operator $\sqrt{\gamma}\s{-}$~\cite{WM10}. We take a fraction $\eta$ of the fluorescence to be observed by homodyne detection, so $\hat b = \sqrt{\gamma\eta}\s{-}$. The remainder is absorbed by the environment, which we model as an unobserved record of photon counts, as discussed above,  with $\hat c = \sqrt{\gamma(1-\eta)}\s{-}$. For a fixed $\both{Y}$ we can compare $\rho\sm$ with $\rho\fil$ on the interval $[0,T]$.  At the final time $\rho\sm(T) = \rho\fil(T)$ because there is no more future record $\fut{Y_{ T}}$ 
to give extra information to $\rho\sm(T)$. Also, we take the initial state to be pure, $\rho_0 = \ket{1}\bra{1}$, which guarantees that  $\rho_{\rm T}$ is pure and that  $\rho\sm(0) = \rho\fil(0)$. 

 To evaluate the advantage gained by smoothing over filtering, we use the purity, 
\begin{equation}\label{eq:purity}
P[\rho\con(t)]=\Tr[\rho\con^2(t)],
\end{equation}
 where $\rho\con$ could be either $\rho\fil$ or $\rho\sm$. If (as  is the case  in simulations) we know the `true' unobserved record $\both{N}^{\rm true}$ we can also calculate the 
fidelity of the conditioned state 
to the 
true state $\rho\god(t) = \rho_{\past{Y_{ t}}, \past{N_{ t}}^{\rm true}}(t)$, 
\beq \label{eq:fidelity}
F[\rho\god(t),\rho\con(t)] = \Tr[\rho\god(t)\, \rho\con(t)].
\eeq
It is easy to show \cite{GueWis15} that these measures are related by 
\beq \label{eq:identity} 
{\rm E}\cu{P[\rho\con(t)]} = {\rm E}\cu{F[\rho_{\rm T}(t), \rho\con(t)]}
\eeq
where the ensemble averages here are over the  actual distributions for  $\both{N}^{\rm true}$ and $\both{Y}$.

\begin{figure}[t!]
     \begin{center}
            \includegraphics[width=0.5\textwidth]{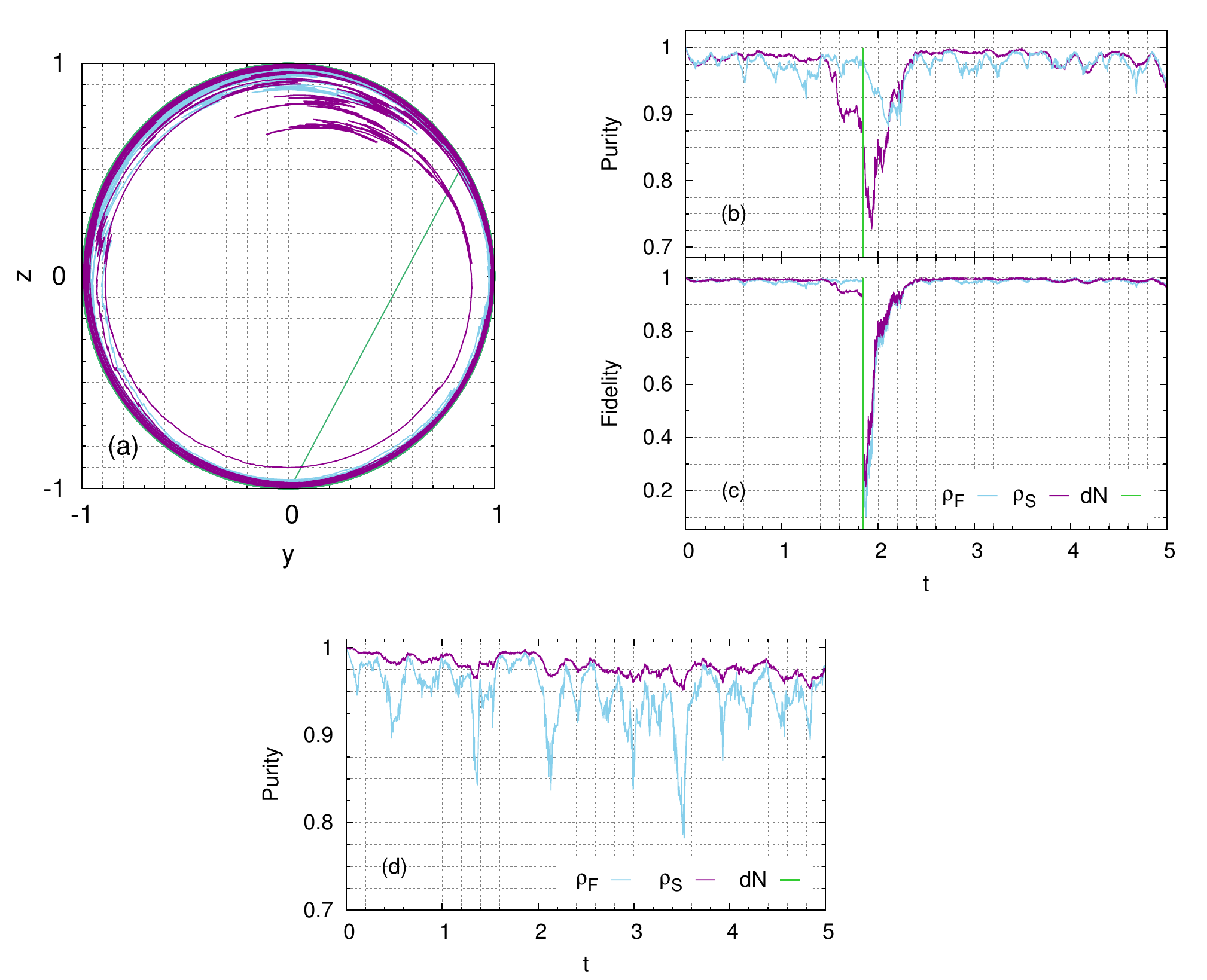}
            
     \vspace{-0.5cm}  
    \end{center} 
    \caption{ [Color online.] (a) Trajectories in the Bloch sphere  for our model system with $\Omega = 20$, $\gamma = 1$, and $\eta = 10/11$. 
    The states shown are $\rho\fil$ (filtered, blue), $\rho\sm$ (smoothed, purple) and $\rho\god$ (`true', green) for a case where the `true' record includes a jump. 
We also plot the purities (b) and fidelities with $\rho\god$  (c) of these $\rho\fil$ and $\rho\sm$. The purities for a record with no jump are shown in (d).  To compute $\rho_{\sm}$ we average over an ensemble of $10^4$ hypothetical unobserved records \unexpanded{$\both{N}$}. 
    }%
   \label{fig:singletraj}
\end{figure}
In Fig.~\ref{fig:singletraj}(a) we show typical trajectories, for $Y$-homodyne [$\Phi=\pi/2$] for a randomly generated true state $\rho\god(t) = \rho_{\past{\nbf Y_{ t}},\past{\nbf N_{t}}}(t)$ featuring one jump at $t\approx 1.8$.  
In this, case, $\rho\god$, $\rho\sm$, and $\rho\fil$ 
are all confined to the $Y$--$Z$ plane of the Bloch sphere, as shown. We plot  \erfs{eq:purity}{eq:fidelity} in Fig.~\ref{fig:singletraj}(b) and (c) respectively. It is notable from (b) that $\rho\sm$ {\em anticipates} the jump in $\rho\god$ and its uncertainty about the timing of the jump leads to a lower purity in the region of the jump than the non-anticipating $\rho\fil$. Similarly, (c) shows that the fidelity of $\rho\sm$ to $\rho\god$ decreases below that of $\rho\fil$ prior to the jump, but is higher after the jump. In (d) we see that if there is no jump, the fidelity with $\rho\god$ is always greater for $\rho\sm$.  

 \begin{figure}[t]
\begin{center}
\includegraphics[scale=.75]{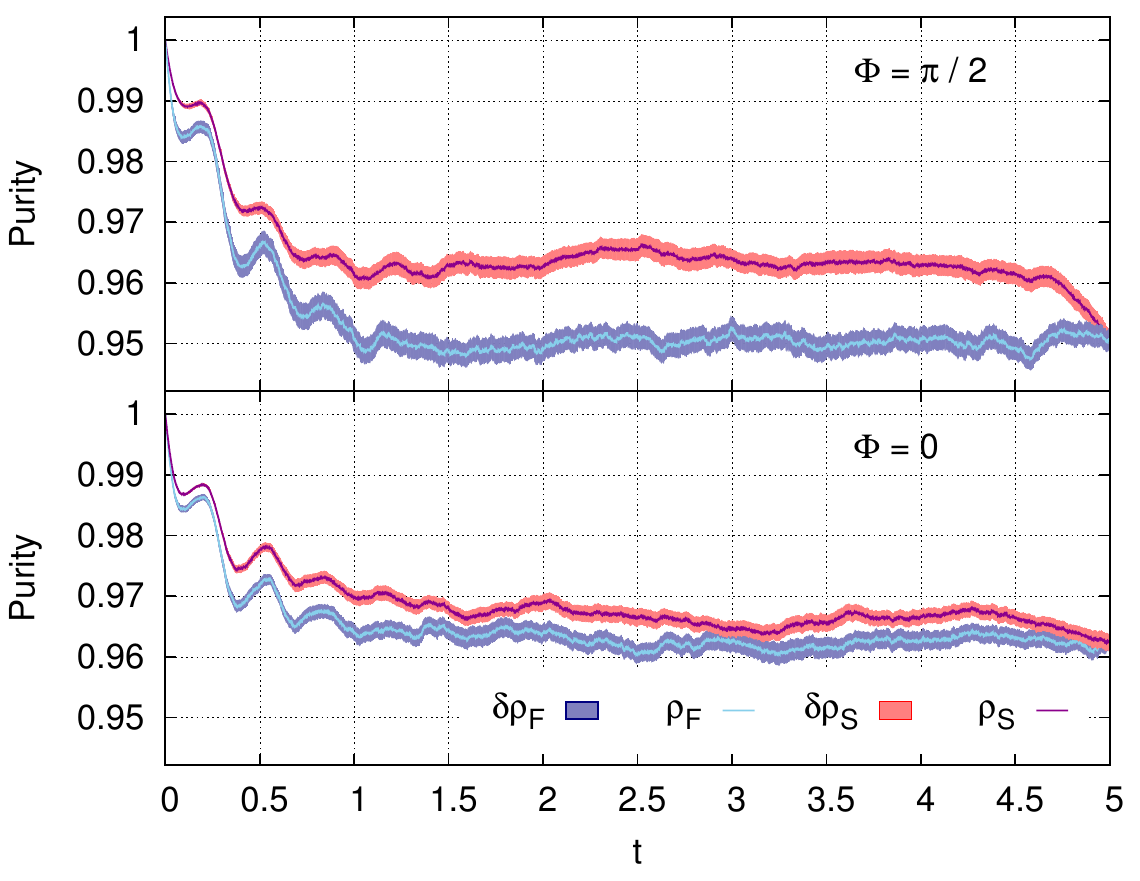}
\caption{Average purity for the case considered for filtered and smoothed states. The average has been calculated with $10^3$ \unexpanded{$\both{Y}$} records, each one of them calculated with $10^4$ estimated \unexpanded{$\both{N}$} records. The figure shows the results obtained form a Y-homodyne measure (top) or X-homodyne (bottom).}
\label{fig:avpurity}
\end{center}
\end{figure}

We confirm that  smoothing enables better state estimation on average by calculating the average purity, 
for $10^3$ realisations of $\both{Y}$. Recall from \erf{eq:identity} that higher purity means higher fidelity
with the true state. We plot this in Fig.~\ref{fig:avpurity} for two different local oscillator phases: 
$\Phi=\pi/2$ ($Y$-homodyne) in (a) and $\Phi=0$ ($X$-homodyne) in (b). Because the driving of the atom causes $\s{y}$ to oscillate at a frequencies $\Omega \gg \gamma$, it is harder to track the state of the atom using $Y$-homodyne detection, and the purity of the filtered state is lower than for $X$-homodyne detection~\cite{AtkBraJacWis05}. It is the former case which shows the greatest improvement in purity under smoothing: about  26\%  of the purity lost, because of the unobserved radiation, is recovered in (a) compared to about  12\%  in (b). 

One can  easily  show~\cite{GueWis15} that our theory of quantum state smoothing includes as a special case 
the HMM  that applies~\cite{Arm09,GJM13,GMKM14} to quantum systems that (unlike our atomic example) have no coherences and so are effectively classical. For genuinely quantum systems there are many questions about state-smoothing to explore, including: what happens if one assumes the unobserved unravelling to be different from photon counting; 
is there a relation between quantum state smoothing and the ``most probable path'' formalism of Refs.~\cite{Cha13,Web14}; does the HMM inevitably emerge in the classical limit, and does quantum smoothing necessarily work best in that limit; and what experiments would show uniquely quantum aspects of quantum state smoothing?

\acknowledgments 

This research was supported by the Australian Research Council Program CE110001027.

\bibliography{references,books}
\newpage~\newpage
\appendix
\includepdf[pages={1}]{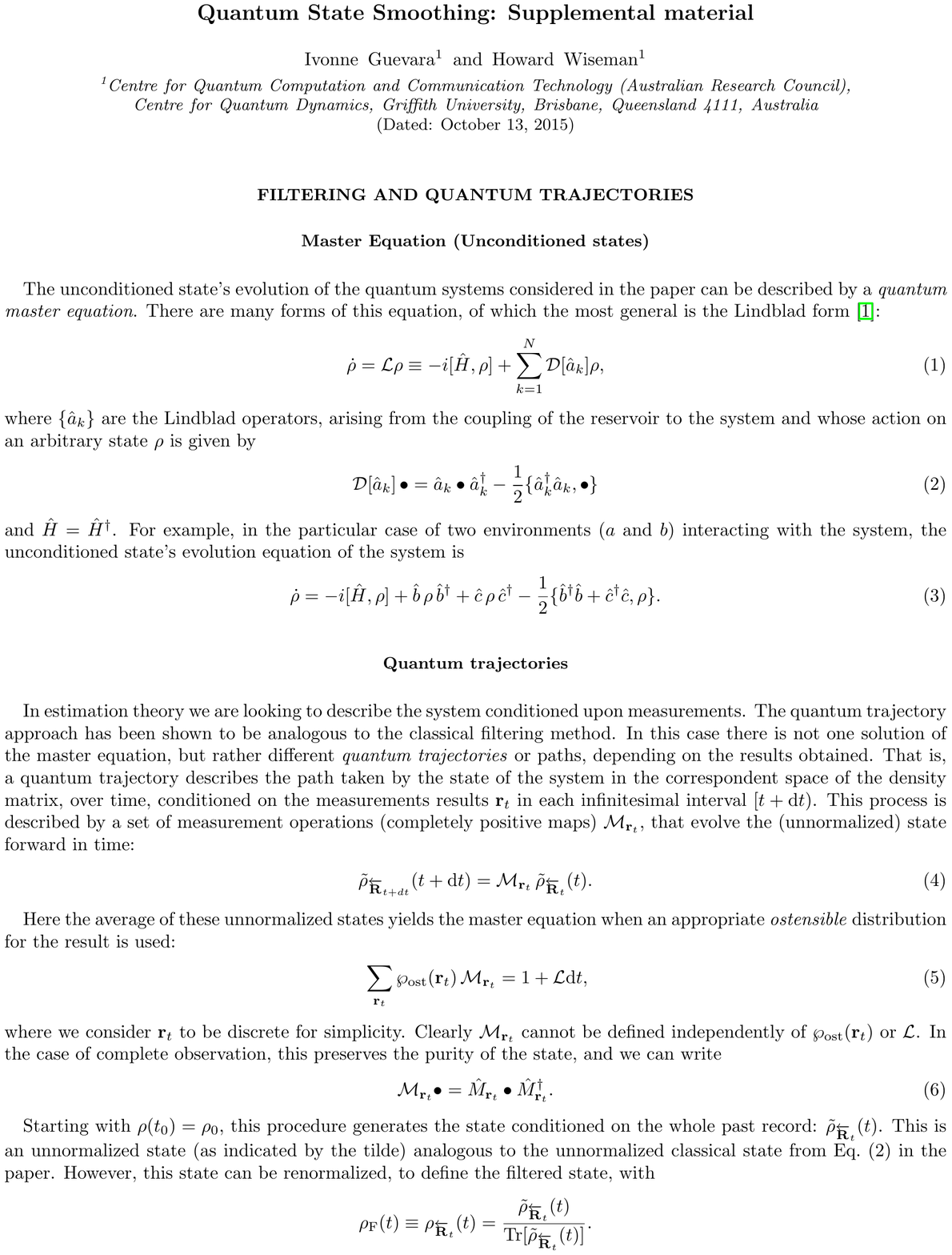}
~\newpage
\includepdf[pages={2}]{supp.pdf}
~\newpage
\includepdf[pages={3}]{supp.pdf}
~\newpage
\includepdf[pages={4}]{supp.pdf}
~\newpage


\section*{Supplementary material}

# Quantum State Smoothing: Supplemental material

Ivonne Guevara[1] and Howard Wiseman[1]

[1]Centre for Quantum Computation and Communication Technology (Australian Research Council),
Centre for Quantum Dynamics, Griffith University, Brisbane, Queensland 4111, Australia

(Dated: October 13, 2015)

## FILTERING AND QUANTUM TRAJECTORIES

### Master Equation (Unconditioned states)

The unconditioned state's evolution of the quantum systems considered in the paper can be described by a *quantum master equation*. There are many forms of this equation, of which the most general is the Lindblad form [1]:

$$\dot{\rho} = \mathcal{L}\rho \equiv -i[\hat{H}, \rho] + \sum_{k=1}^{N} \mathcal{D}[\hat{a}_k]\rho, \qquad (1)$$

where $\{\hat{a}_k\}$ are the Lindblad operators, arising from the coupling of the reservoir to the system and whose action on an arbitrary state $\rho$ is given by

$$\mathcal{D}[\hat{a}_k] \bullet = \hat{a}_k \bullet \hat{a}_k^\dagger - \frac{1}{2}\{\hat{a}_k^\dagger \hat{a}_k, \bullet\} \qquad (2)$$

and $\hat{H} = \hat{H}^\dagger$. For example, in the particular case of two environments ($a$ and $b$) interacting with the system, the unconditioned state's evolution equation of the system is

$$\dot{\rho} = -i[\hat{H}, \rho] + \hat{b}\,\rho\,\hat{b}^\dagger + \hat{c}\,\rho\,\hat{c}^\dagger - \frac{1}{2}\{\hat{b}^\dagger\hat{b} + \hat{c}^\dagger\hat{c}, \rho\}. \qquad (3)$$

### Quantum trajectories

In estimation theory we are looking to describe the system conditioned upon measurements. The quantum trajectory approach has been shown to be analogous to the classical filtering method. In this case there is not one solution of the master equation, but rather different *quantum trajectories* or paths, depending on the results obtained. That is, a quantum trajectory describes the path taken by the state of the system in the correspondent space of the density matrix, over time, conditioned on the measurements results $\mathbf{r}_t$ in each infinitesimal interval $[t + dt)$. This process is described by a set of measurement operations (completely positive maps) $\mathcal{M}_{\mathbf{r}_t}$, that evolve the (unnormalized) state forward in time:

$$\tilde{\rho}_{\overleftarrow{\mathbf{R}}_{t+dt}}(t + dt) = \mathcal{M}_{\mathbf{r}_t}\,\tilde{\rho}_{\overleftarrow{\mathbf{R}}_t}(t). \qquad (4)$$

Here the average of these unnormalized states yields the master equation when an appropriate *ostensible* distribution for the result is used:

$$\sum_{\mathbf{r}_t} \wp_{\text{ost}}(\mathbf{r}_t)\,\mathcal{M}_{\mathbf{r}_t} = 1 + \mathcal{L}dt, \qquad (5)$$

where we consider $\mathbf{r}_t$ to be discrete for simplicity. Clearly $\mathcal{M}_{\mathbf{r}_t}$ cannot be defined independently of $\wp_{\text{ost}}(\mathbf{r}_t)$ or $\mathcal{L}$. In the case of complete observation, this preserves the purity of the state, and we can write

$$\mathcal{M}_{\mathbf{r}_t}\bullet = \hat{M}_{\mathbf{r}_t} \bullet \hat{M}_{\mathbf{r}_t}^\dagger. \qquad (6)$$

Starting with $\rho(t_0) = \rho_0$, this procedure generates the state conditioned on the whole past record: $\tilde{\rho}_{\overleftarrow{\mathbf{R}}_t}(t)$. This is an unnormalized state (as indicated by the tilde) analogous to the unnormalized classical state from Eq. (2) in the paper. However, this state can be renormalized, to define the filtered state, with

$$\rho_{\text{F}}(t) \equiv \rho_{\overleftarrow{\mathbf{R}}_t}(t) = \frac{\tilde{\rho}_{\overleftarrow{\mathbf{R}}_t}(t)}{\text{Tr}[\tilde{\rho}_{\overleftarrow{\mathbf{R}}_t}(t)]}.$$

The unnormalized state carries information about the actual probability distribution of the results measurements as follows

$$\wp(\overleftarrow{\mathbf{R}}_t|\rho_0) = \text{Tr}[\tilde{\rho}_{\overleftarrow{\mathbf{R}}_t}(t)]\,\wp_{\text{ost}}(\overleftarrow{\mathbf{R}}_t|\rho_0). \tag{7}$$

The quantum trajectories can be discontinuous (quantum jumps) or continuous (quantum diffusive trajectories) depending on the kind of measurements that are conditioning the evolution.

### Diffusive Quantum Trajectories

We consider now the case where the measurements are diffusive in nature and the record in each interval is a continuous variable. For example, for the case of unit-efficiency homodyne detection with a single channel $b$, the result $y_t$ is a real number and we can choose $\wp_{\text{ost}}(y_t)$ to be a zero-mean Gaussian of variance $1/dt$. Then the measurement operation is defined by $\mathcal{M}_{y_t} = \hat{M}_{y_t} \bullet \hat{M}_{y_t}^\dagger$, where [2]

$$\hat{M}_{y_t} = 1 - i\hat{H}dt - \frac{1}{2}\hat{b}^\dagger\hat{b}\,dt + e^{-i\Phi}\hat{b}\,y_t dt, \tag{8}$$

is the measurement operator acting only in the system Hilbert space. It is easy to verify that this reproduces the master equation evolution to order $dt$:

$$\int dy_t\,\wp_{\text{ost}}(y_t)\,\mathcal{M}_{\mathbf{y}_t}\bullet = 1 - i[\hat{H},\bullet] + \mathcal{D}[\hat{b}]\bullet. \tag{9}$$

Here $\Phi$ is the phase of the local oscillator. We say we make an $X$-homodyne measurement when $\Phi = 0$ and $Y$-homodyne when $\Phi = \pi/2$.

A normalised version of the filtering state can be expanded using Itô calculus, yielding

$$\rho_{\overleftarrow{Y}_{t+dt}}(t+dt) = \rho_{\overleftarrow{Y}_t}(t) + \left(-i[\hat{H},\rho_{\overleftarrow{Y}_t}(t)] + \mathcal{D}[\hat{b}]\,\rho_{\overleftarrow{Y}_t}(t)\right)dt + \left(e^{i\Phi}(\hat{b}-\langle\hat{b}\rangle)\rho_{\overleftarrow{Y}_t}(t) + \text{H.c.}\right)dw(t). \tag{10}$$

Since this is a normalized state, it averages to the master equation solution according to the *actual* probability distribution, for which $dw(t) \equiv y_t dt - \text{Tr}[e^{i\Phi}\hat{b} + e^{-i\Phi}\hat{b}^\dagger]dt$ is a Wiener increment satisfying $\mathrm{E}_{\mathbf{y}_t}[dw(t)] = 0$ and $\mathrm{E}_{\mathbf{y}_t}[dw^2(t)] = dt$. In this form it is elementary to see that averaging over the measurement results yields the master equation again. We can use this equation to directly calculate the normalised $\rho_F = \rho_{\overleftarrow{Y}_t}$.

### Discrete Quantum Trajectories (Quantum Jumps)

A similar definition, or the case of a single-channel with unit-efficiency photon counting:

$$\mathcal{M}_{n_t}\bullet = \hat{M}_{n_t}\bullet\hat{M}_{n_t}^\dagger.$$

In this case the record is discrete — there are only two possible results in the interval $[t, t+dt]$: one photon or none. Choosing ostensible probabilities

$$\wp_{\text{ost}}(n_t := 1) = \lambda dt\,;\quad \wp_{\text{ost}}(n_t := 0) = 1 - \lambda dt, \tag{11}$$

the correspondent measurement operators are [2]

$$\begin{aligned}\hat{M}_1 &= \hat{c}/\sqrt{\lambda},\\ \hat{M}_0 &= 1 - \left(i\hat{H} + \tfrac{1}{2}(\hat{c}^\dagger\hat{c} - \lambda)\right)dt.\end{aligned} \tag{12}$$

Again, it can easily be verified that averaging by using these ostensible probabilities gives

$$\begin{aligned}\sum_{n_t}\wp_{\text{ost}}(n_t)\,\tilde{\rho}_{n_t}(t) &= (1-\lambda dt)\hat{M}_0\,\rho(t)\hat{M}_0^\dagger + \lambda dt\hat{M}_1\,\rho(t)\hat{M}_1^\dagger\\ &= \rho(t) - i[\hat{H},\rho(t)]dt + \mathcal{D}[\hat{c}]\rho(t)dt\end{aligned} \tag{13}$$

as required for consistency with the master equation. The *actual* probabilities, which relate to the normalized state, are determined by

$$\wp(n_t := 1) = \wp_{\text{ost}}(n_t := 1)\text{Tr}[\hat{M}_1\rho(t)\hat{M}_1^\dagger] = \text{Tr}[\rho(t)\hat{c}^\dagger\hat{c}]dt, \tag{14}$$

and similarly for $\wp(n_t := 0)$, which evaluates to $1 - \wp(n_t = 1)$.

### Multiple channel monitoring

When the system is monitored through more than one channel, as we must consider for the case of the quantum smoothing state method, the conditioned state is defined in terms of the measurement operators of both channels,

$$\mathcal{M}_{n_t,y_t}\bullet = M_{y_t} M_{n_t} \bullet M_{n_t}^\dagger M_{\mathbf{y}_t}^\dagger. \tag{15}$$

This operation gives Eq. (3) when averaging over the results for both monitorings if we take the Hamiltonian in each measurement operator to be $\hat{H}/2$ rather than the full $\hat{H}$. In the case of average over diffusive trajectories, we have to order $\mathrm{d}t$,

$$\begin{aligned}\int \mathrm{d}y_t\, \wp_{\mathrm{ost}}(y_t)\mathcal{M}_{n_t,y_t}\bullet &= \int \mathrm{d}y_t\, \wp_{\mathrm{ost}}(y_t)\, \hat{M}_{y_t}\, (\hat{M}_{n_t} \bullet \hat{M}_{n_t}^\dagger)\hat{M}_{y_t}^\dagger \\ &= \hat{M}_{n_t} \bullet \hat{M}_{n_t}^\dagger - \tfrac{i}{2}[\hat{H},\bullet]\mathrm{d}t + \mathcal{D}[\hat{b}]\bullet \mathrm{d}t \\ &\equiv \mathcal{M}_{n_t}\bullet. \end{aligned} \tag{16}$$

and similarly for discrete measurements,

$$\begin{aligned}\sum_{n_t} \wp_{\mathrm{ost}}(\mathbf{n}_t)\, \mathcal{M}_{n_t,y_t}\bullet &= \sum_{n_t} \wp_{\mathrm{ost}}(\mathbf{n}_t)\, \hat{M}_{y_t}\, \hat{M}_{n_t} \bullet \hat{M}_{n_t}^\dagger\, \hat{M}_{\mathbf{y}_t}^\dagger, \\ &= \hat{M}_{y_t} \bullet \hat{M}_{y_t}^\dagger - \tfrac{i}{2}[\hat{H},\bullet]\mathrm{d}t + \mathcal{D}[\hat{c}]\bullet \mathrm{d}t \\ &\equiv \mathcal{M}_{y_t}\bullet. \end{aligned} \tag{17}$$

Just as for the individual jump and diffusion case, we can easily normalize the state, and generate records with the correct actual distribution.

### Choosing the Ostensible probabilities

The ostensible probability distribution for the measurement results arbitrary, as long as it is fixed for a given ensemble for simulation purposes. Thus, for the purpose of smoothing, where we have to simulate many different unobserved records $\overleftrightarrow{\mathbf{U}}$ for a given observed record $\overleftrightarrow{\mathbf{O}}$, the ostensible distribution for $\overleftrightarrow{\mathbf{U}}$ could depend upon $\overleftrightarrow{\mathbf{O}}$ (since this is fixed). For the most efficient simulations (that is, giving the smallest statistical errors for a given ensemble size), we would like the ostensible distribution $\wp^{\mathrm{ost}}(\overleftrightarrow{\mathbf{U}})$ to be as close as possible to the actual distribution $\wp(\overleftrightarrow{\mathbf{U}}|\overleftrightarrow{\mathbf{O}})$. This last is, however, exactly what we are trying to determine from the simulations — the smoothed distribution $\wp_{\mathrm{S}}(\overleftrightarrow{\mathbf{U}})$ — so we cannot expect to just guess it. Moreover, the more complicated $\wp^{\mathrm{ost}}(\overleftrightarrow{\mathbf{U}})$ is, the more complicated the quantum trajectory equations corresponding to this ostensible distribution will be. Thus, for simplicity, we choose an ostensible probability distribution using only the past information $\overleftarrow{\mathbf{O}}$, giving a distribution that factorizes at different times. For the example in the paper, where $O = Y$ and $U = N$, the distribution for $n_t$ is thus described by a parameter $\lambda$ as introduced in Eq. (11), but this becomes time-dependent and determined by the filtered state $\rho_{\mathrm{F}}(t) = \rho_{\overleftarrow{Y}_t}(t)$:

$$\lambda(t) = \mathrm{Tr}[\hat{c}^\dagger \hat{c}\, \rho_{\mathrm{F}}(t)]. \tag{18}$$

That is, we choose the ostensible rate of unobserved jumps (photon counts) to equal the rate that would pertain, as in Eq. (14), if $\rho_{\mathrm{F}}(t)$ were the state of the system (which it is not — the unknown $\rho_{\overleftarrow{Y}_t,\overleftarrow{N}_t}(t)$ is the 'true' state $\rho_{\mathrm{T}}(t)$). A more sophisticated approach, which may give somewhat smaller standard deviation, would be to use a small ensemble to generate an approximation to $\rho_{\mathrm{S}}(t)$, and then use this, in place of $\rho_{\mathrm{F}}(t)$, in Eq. (18), to generate a new approximation to $\rho_{\mathrm{S}}(t)$ from a larger ensemble of unknown records.

Note that the use of ostensible probability distributions is not limited to quantum systems. For example, the classical analogue of Eq. (7) is

$$\wp(\overleftarrow{\mathbf{R}}_t|\wp_0) = \left[\int \mathrm{d}\mathbf{x}'\, \tilde{\wp}_{\overleftarrow{\mathbf{R}}_t}(\mathbf{x}'_t)\right]\, \wp_{\mathrm{ost}}(\overleftarrow{\mathbf{R}}_t|\wp_0). \tag{19}$$

## SMOOTHING PROBABILITY

In this section we will show how to manipulate the expressions relating to probabilities to find the smoothed probability $\wp_S(\overleftarrow{\mathbf{U}}_t) = \wp_{\overleftrightarrow{\mathbf{O}}}(\overleftarrow{\mathbf{U}}_t) = \Pr[\overleftarrow{\mathbf{U}}_t^{\text{true}} = \overleftarrow{\mathbf{U}}_t | \overleftrightarrow{\mathbf{O}}, \rho(t_0)]$. Starting from the probability definition

$$\wp_S(\overleftarrow{\mathbf{U}}_t) = \wp(\overleftarrow{\mathbf{U}}_t | \overrightarrow{\mathbf{O}}_t, \overleftarrow{\mathbf{O}}_t)$$

$$= \frac{\wp(\overleftarrow{\mathbf{U}}_t | \overrightarrow{\mathbf{O}}_t, \overleftarrow{\mathbf{O}}_t) \wp(\overrightarrow{\mathbf{O}}_t, \overleftarrow{\mathbf{O}}_t)}{\wp(\overleftarrow{\mathbf{U}}_t, \overleftarrow{\mathbf{O}}_t)} \frac{\wp(\overleftarrow{\mathbf{U}}_t, \overleftarrow{\mathbf{O}}_t)}{\wp(\overrightarrow{\mathbf{O}}_t, \overleftarrow{\mathbf{O}}_t)},$$

we can use Baye's rule to rewrite the expression as:

$$\wp_S(\overleftarrow{\mathbf{U}}_t) = \frac{\wp(\overrightarrow{\mathbf{O}}_t | \overleftarrow{\mathbf{U}}_t, \overleftarrow{\mathbf{O}}_t) \wp(\overleftarrow{\mathbf{U}}_t, \overleftarrow{\mathbf{O}}_t)}{\wp(\overrightarrow{\mathbf{O}}_t, \overleftarrow{\mathbf{O}}_t)},$$

so that, as $\wp(A, B) = \wp(A|B)\wp(B)$ and $\wp(A) = \sum_B \wp(A|B)\wp(B)$, we finally get,

$$\wp_S(\overleftarrow{\mathbf{U}}_t) = \frac{\wp(\overrightarrow{\mathbf{O}}_t | \overleftarrow{\mathbf{U}}_t, \overleftarrow{\mathbf{O}}_t) \wp(\overleftarrow{\mathbf{U}}_t | \overleftarrow{\mathbf{O}}_t)}{\sum_{\overleftarrow{\mathbf{U}}_t} \wp(\overrightarrow{\mathbf{O}}_t | \overleftarrow{\mathbf{U}}_t, \overleftarrow{\mathbf{O}}_t) \wp(\overleftarrow{\mathbf{U}}_t | \overleftarrow{\mathbf{O}}_t)}. \tag{20}$$

## AVERAGE PURITY AND AVERAGE FIDELITY

The smoothed quantum state's purity, and its fidelity with the true state, are appropriate measures of the model's success. We will demonstrate in this section that for the case where the 'true' state is pure, the values of these two quantities, averaged over all records (observed and unobserved), is identical. This is true both for filtering and smoothing.

In Eq. (9) in the paper, the quantum smoothed state is defined as an ensemble average over realizations of the unknown record $\overleftarrow{\mathbf{U}}_t$. If we define a general conditioning C, filtered and smoothed states can be described similarly,

$$\rho_C(t) = \mathrm{E}_{\overleftarrow{\mathbf{U}}_t | \mathbf{O}_\Omega}[\rho_{\overleftarrow{\mathbf{O}}_t, \overleftarrow{\mathbf{U}}_t}(t)] \equiv \sum_{\overleftarrow{\mathbf{U}}} \wp_C(\overleftarrow{\mathbf{U}}_t) \rho_{\overleftarrow{\mathbf{O}}_t, \overleftarrow{\mathbf{U}}_t}(t), \tag{21}$$

where $\mathbf{O}_\Omega = \overleftrightarrow{\mathbf{O}}$ for smoothing (C=S), and $\mathbf{O}_\Omega = \overleftarrow{\mathbf{O}}_t$ for filtering (C=F). From the definition for purity in Eq.(13)

$$\begin{aligned} P[\rho_C(t)] &= \mathrm{Tr}\left[\rho_C(t) \rho_C(t)\right] \\ &= \mathrm{Tr}\left[\sum_{\overleftarrow{\mathbf{U}}_t} \wp_C(\overleftarrow{\mathbf{U}}_t) \rho_{\overleftarrow{\mathbf{O}}_t, \overleftarrow{\mathbf{U}}_t}(t) \rho_C(t)\right] \\ &= \sum_{\overleftarrow{\mathbf{U}}_t} \wp_C(\overleftarrow{\mathbf{U}}_t) \mathrm{Tr}[\rho_{\overleftarrow{\mathbf{O}}_t, \overleftarrow{\mathbf{U}}_t}(t) \rho_C(t)]. \end{aligned} \tag{22}$$

Also, when the true state $\rho_T$ is pure,

$$\begin{aligned} F[\rho_C(t), \rho_T] &= \mathrm{Tr}\left[\rho_C(t) \rho_T(t)\right] \\ &= \mathrm{Tr}\left[\rho_{\overleftarrow{\mathbf{O}}_t, \overleftarrow{\mathbf{U}}_t^{\text{true}}}(t) \rho_C(t)\right]. \end{aligned} \tag{23}$$

Averaging Eq. (22) over the appropriate ensemble of measurement records $\mathbf{O}_\Omega$,

$$\begin{aligned} \mathrm{E}_{\mathbf{O}_\Omega}\{P[\rho_C(t)]\} &= \sum_{\mathbf{O}_\Omega} \wp(\mathbf{O}_\Omega) \sum_{\overleftarrow{\mathbf{U}}_t} \wp(\overleftarrow{\mathbf{U}}_t | \mathbf{O}_\Omega) \mathrm{Tr}[\rho_{\overleftarrow{\mathbf{O}}_t, \overleftarrow{\mathbf{U}}_t}(t) \rho_C(t)] \\ &= \sum_{\mathbf{O}_\Omega} \sum_{\overleftarrow{\mathbf{U}}_t} \wp(\overleftarrow{\mathbf{U}}_t | \mathbf{O}_\Omega) \wp(\mathbf{O}_\Omega) \mathrm{Tr}[\rho_{\overleftarrow{\mathbf{O}}_t, \overleftarrow{\mathbf{U}}_t}(t) \rho_C(t)] \\ &= \sum_{\mathbf{O}_\Omega, \overleftarrow{\mathbf{U}}_t} \wp(\mathbf{O}_\Omega, \overleftarrow{\mathbf{O}}_t) \mathrm{Tr}\left[\rho_{\overleftarrow{\mathbf{O}}_t, \overleftarrow{\mathbf{U}}_t}(t) \rho_C(t)\right] \\ &= \mathrm{E}_{\mathbf{O}_\Omega, \overleftarrow{\mathbf{U}}_t}\left\{\mathrm{Tr}[\rho_{\overleftarrow{\mathbf{O}}_t, \overleftarrow{\mathbf{O}}_t}(t) \rho_C(t)]\right\} \\ &= \mathrm{E}_{\mathbf{O}_\Omega, \overleftarrow{\mathbf{U}}_t}\left\{F[\rho_T(t), \rho_C(t)]\right\}, \end{aligned} \tag{24}$$

where here we can think of the ensemble average as being over all possible 'true' records $\overleftrightarrow{\mathbf{U}}$ as well as $\overleftrightarrow{\mathbf{O}}$.

# RELATION TO CLASSICAL HIDDEN MARKOV MODELS

In the paper we mention that the results relating to HMMs in [3, 4], and also in the prior work of [5] (which is actually somewhat more general), can be easily obtained as a restriction of the quantum state smoothing model. In these systems, the system dynamics—the quantum master equation—can be reformulated, using a suitable basis, as a classical master equation. That is, the stochastic dynamics can be described as transitions (quantum jumps) between *orthogonal* states of the system. This means that, independently of whatever record $\overleftrightarrow{\mathbf{O}}$ an observer Alice has, a second, hypothetical observer (Bob) can always know exactly what state the system is in. That is, Bob's unobserved (by Alice) jump record $\overleftarrow{N}_t$ completely determines the state of the system at time $t$, as the rank-one projector $\hat{\Pi}_{x(\overleftarrow{N}_t)}$. The label $x$, identifying which state the system is in at time $t$, is the HMM for the system.

We will now show in detail the connection between our quantum state smoothing technique and the HMM technique. Recall our method for calculating the quantum smoothed state:

$$\rho_{\mathrm{S}}(t) = \sum_{\overleftarrow{N}_t} \wp_{\mathrm{S}}(\overleftarrow{N}_t) \rho_{\overleftrightarrow{\mathbf{O}}_t, \overleftarrow{N}_t} \tag{25}$$

where $\wp_{\mathrm{S}}(\overleftarrow{N}_t) \equiv \wp(\overleftarrow{N}_t | \overleftrightarrow{\mathbf{O}})$. Now in this case $\rho_{\overleftrightarrow{\mathbf{O}}_t, \overleftarrow{N}_t} = \hat{\Pi}_{x(\overleftarrow{N}_t)}$. Introducing the completeness relation $\sum_{x'} \delta_{x',x} = 1$, for any value of the classical variable $x$, and also using

$$\delta_{x',x(\overleftarrow{N}_t)} = \wp(x' | \overleftrightarrow{\mathbf{O}}, \overleftarrow{N}_t),$$

we can rewrite the smoothed state as

$$\rho_{\mathrm{S}}(t) = \sum_{x'} \sum_{\overleftarrow{N}_t} \wp(x' | \overleftrightarrow{\mathbf{O}}, \overleftarrow{N}_t) \wp(\overleftarrow{N}_t | \overleftrightarrow{\mathbf{O}}) \hat{\Pi}_{x'} \tag{26}$$

$$= \sum_{x'} \wp(x' | \overleftrightarrow{\mathbf{O}}) \hat{\Pi}_{x'} = \sum_{x} \wp_{\mathrm{S}}(x) \hat{\Pi}_{x'}. \tag{27}$$

That is, the problem reduces to the classical smoothing problem, with the simple solution (requiring no ensemble averaging)

$$\wp_{\mathrm{S}}(x) \propto \wp(\overrightarrow{\mathbf{O}} | x) \wp(x | \overleftarrow{\mathbf{O}}_t),$$

as used by Armen *et al.* [5] and Gammelmark *et al.* [3, 4].

---



\end{document}